\documentclass{ws-ijmpd}
\usepackage[super,compress]{cite}
\usepackage{graphicx}

\begin{document}

\markboth{Authors' Names}
{Instructions for Typing Manuscripts (Paper's Title)}

%
\catchline{}{}{}{}{}
%

\title{DYNAMICS OF A PARTICLE IN THE GENERALISED ELLIS-BRONNIKOV WORMHOLE ON THE ROTATING ARCHIMEDE'S SPIRAL}

\author{N. GIORGADZE}

\address{School of Physics, Free University of Tbilisi\\
Tbilisi, 0159, Georgia\\
ngior17@freeuni.edu.ge}

\author{Z.N. OSMANOV}

\address{School of Physics, Free University of Tbilisi\\
Tbilisi, 0159, Georgia\\
Kharadze Georgian National Astrophysical Observatory\\
Abastumani, 0301, Georgia\\
z.osmanov@freeuni.edu.ge}

\maketitle

\begin{history}
\received{Day Month Year}
\revised{Day Month Year}
\end{history}

\begin{abstract}
A particle moving on curved magnetic field lines in the wormhole metrics is considered. The dynamics of a single particle is studied on a co-rotating parameterized magnetic field line. Being interested in the force-free dynamics we choose the rotating Archimedes' spiral embedded in the generalized Ellis-Bronnikov metrics. 
By analysing the phase space, we explore how the physical parameters affect the motion of the particle and it is shown that the particles may reach the force free regime, eventually escaping the wormhole.
\end{abstract}


\ccode{PACS numbers:}


\section{Introduction}	 

Existence of wormholes (WH) has frequently been discussed in the scientific community.  It is argued that WHs might be created during the inflation period of the early evolution of the universe (Refs. \refcite{Wheeler1}-\refcite{Hawking}). It is also assumed that there is accreting matter in the central region of WHs (Ref. \refcite{Harko}). Moreover It has been considered that some astrophysical objects may be or had been entrances to WHs (Ref. \refcite{Kardashev}). Such astrophysical objects not only may provide magnetic field of extreme strength of order $10^{13}$ G, (Ref. \refcite{Kardashev}), but often they also are characterized by rotation, therefore, on the one hand extremely strong magnetic field may guarantee the frozen-in condition and on the other hand, particles will slide along field lines, experiencing relativistic centrifugal force. 

The motion of matter may be studied either as fluid dynamics of an ensemble of particles (Ref. \refcite{Butb}) or as dynamics of a single particle (Refs. \refcite{Ars,Bakh,Gud,AR  MG,AR GD ZO}). In this paper we will be examining motion of a charged particle in the presence of magnetic field. As we have already mentioned, the super-strong magnetic field will result in the frozen-in condition of a particle on "field lines", causing it to slide across them. In particular, one can straightforwardly estimate that the gyroradius of a proton $\gamma mc^2/eB$ with the relativistic factor $\gamma\simeq 10^7$ is of the order of $1$ cm for $B\simeq 10^{13}$ Gauss, which for electrons is even less. Taking rotation of the magnetic field lines into account holds particularly interesting results as particle is subject to the relativistic centrifugal acceleration.

Dynamics of particles on rotating channels has been studied in a series of papers (Refs. \refcite{Ars}-\refcite{ZO FR}). Problem becomes interesting near the light cylinder (LC) region, where the velocity of rotation coincides with the speed of light. In a simplified model considering the motion of the particle in rectilinear rotating tube it has been shown that near the LC zone the particle changes the sign of the radial acceleration (Ref. \refcite{AR MG}). This is not a realistic behaviour because in the the study the maintainance of the rigid rotation has been assumed. However, in real astrophysical situations field lines are curvilinear as they tend to lag behind the rotation near the LC surface. We take special interest in field lines having the configuration of the Archimede's spiral, because, as it has been shown, the force-free regime may be established, enabling particles to reach infinity with finite velocity (Refs. \refcite{Ars,Bakh}).

In (Ref.~\refcite{Ars}) the simplest form of WH metrics proposed by Ellis (Refs. \refcite{Ellis}) and Bronnikov (Refs. \refcite{Bronnikov}) was considered. In the most simplified Ellis-Bronnikov (EB) WH metrics rotating Archimede's spiral was embedded and it was shown that particles can escape the WH with finite velocity. In this paper we will be considering motion of the particle on rotating spiral in the generalized Ellis-Bronnikov (GEB) metrics showing that the particles will still reach the force-free regime and escape the WH.

The outline of the paper will be following: In section 2 we consider a theoretical model for motion of the particle on any curve in any given metrics; in section 3 we reduce our problem to GEB wormhole and Archimede's spiral and consider results.

\section{Main Consideration}

Let us first consider, without specifying dimensions and characteristics of the space, a most general form of metrics  

\begin{equation}
	\textit{$ds^2 = g_{\alpha \beta}dq^\alpha dq^\beta$}
	\label{eqn1}
\end{equation}

with the metric tensor components \textit{$g_{\alpha \beta} = g_{\alpha\beta}(q^\mu)$} being functions of the coordinates in the laboratory frame of reference \textit{$q^\mu$}.

Lagrangian of the particle moving in such space is given by 
\begin{equation}
	L =- \frac{1}{2}g_{\alpha \beta}\dot{q}^\alpha\dot{q}^\beta,
	\label{eqn2}
\end{equation}
where $\dot{q}^\mu \equiv \frac{dq^\mu}{d\tau}$ and $\tau$ - a parameter of evolution, is the proper time of the particle.


From the Euler-Lagrange equations
\begin{equation}  
	\frac{\partial L}{\partial q^\alpha} = \frac{d}{d\tau}\biggl(\frac{\partial L}{\partial \dot{q}^\alpha}\biggl),
	\label{eqn3}
\end{equation}

one can straightforwardly obtain the well known geodesic equations of motion:
\begin{equation}
	\ddot{q}^\beta = -\Gamma^\beta_{\mu\nu} \dot{q}^\mu\dot{q}^\nu
	\label{eqn4}
\end{equation}
with \textit{$\Gamma^\beta_{\mu\nu} = \frac{1}{2}g^{\alpha\beta}\bigr[\partial_\mu g_{\alpha\nu} + \partial_\nu g_{\alpha\mu} - \partial_\alpha g_{\mu\nu}\bigl]$}.



Note that here we have not yet made any assumptions about the space or introduced any constraints on the particle. As already mentioned above, we will be discussing motion of the particle on rotating field lines. We seek to constrain a particle on a certain field line, i.e. to constrain it on one-dimensional curve embedded in three-dimensional space. In other words we aim to describe the motion in $\textit{1+1}$ rather than $\textit{1+3}$ formalism. Having Parameterized the motion alongside the field line, equation of motion expressed by Eq. (\ref{eqn4}) would carry the idea that particle moves on geodesic lines in derived $\textit{1+1}$ metrics.

Parameterization of a field line is achieved through expressing all the coordinates as a function of lab-time and one spatial coordinate we wish to examine. 

\begin{equation}
	q^2 = q^2(q^0, q^1)  \qquad \text{and} \qquad q^3 = q^3(q^0, q^1),
	\label{eqn5}
\end{equation}
reducing the metric tensor to the following form
\begin{equation}
	ds^2 =  \tilde{g}_{\alpha \beta}dq^\alpha dq^\beta =  \tilde{g}_{00}(dq^0)^2 + 2\tilde{g}_{01}dq^0dq^1 + \tilde{g}_{11}(dq^1)^2.
	\label{eqn6}
\end{equation}
Here $\tilde{g}_{\alpha \beta}$ is given by the general transformation rule of tensors when changing the coordinate system and henceforth, we will be referring to it as $g_{\alpha \beta}$. 
From Eq. (\ref{eqn3}), through the straightforward calculations we obtain the equation of motion for one remaining spatial coordinate

\begin{equation}
		\frac{d^2q^1}{dt^2} = \frac{1}{2\Delta g}\bigr[A_0 
		+ A_1 v 
		+A_2 v^2 
		+ A_3 v^3
		\bigl],
		\label{eqn7}
\end{equation}
where the coefficients $A_i$ are the functions of metrics and depend only on the spatial coordinates
\begin{equation}
	A_0 = g_{00}\partial_{1}g_{00},
	\label{eqn8}
\end{equation}
\begin{equation}
	A_1 = 3g_{01}\partial_{1} g_{00},
	\label{eqn9}
\end{equation}
\begin{equation}
	A_2 = 2g_{01}\partial_{1} g_{01} + g_{11}\partial_{1} g_{00} - g_{00}\partial_{1} g_{11},
	\label{eqn10}
\end{equation}
\begin{equation}
	A_3 = g_{11}\partial_{1} g_{01} - g_{01}\partial_{1} g_{11},
	\label{eqn11}
\end{equation}
and $\Delta g$ denotes the determinant of the derived metric tensor and $v \equiv \frac{dq^1}{dt}$ is the velocity of the particle alongside the field line as it is observed from the laboratory frame of reference. Deriving the acceleration as a function of coordinates and velocities enables us to further examine motion of the particle that will be explored in next section.

We can examine the dynamics of a particle by finding the conserved quantity of the Lagrangian given by Eq. (\ref{eqn2}). Since the latter does not depend on time the corresponding quantity, associated to the energy of the particle writes as
\begin{equation}
	E = -\gamma (g_{00} + g_{01}v) = const
	\label{eqn12}
\end{equation}
where by definition the Lorentz factor is given by
\begin{equation}
	\gamma = (-g_{00} - 2g_{01}v - g_{11}v^2)^{\frac{1}{2}},
	\label{eqn13}
\end{equation}
Solving Eqs. (\ref{eqn12},\ref{eqn13}) with respect to $v$ one obtains:

\begin{equation}
	v = \frac{\sqrt{g_{00} + E^2}}{g_{01}^2 + E^2 g_{11}}\biggl [-g_{01}\sqrt{g_{00} + E^2} \pm E\sqrt{g_{01}^2 -g_{00}g_{11}} \biggr],
	\label{eqn14}
\end{equation}
which formally is the similar expression derived in (Refs. \refcite{Gud,AR GD ZO})).

\section{Generalized Ellis-Bronnikov Wormhole}

In this section we consider motion of the particle on rotating field lines in a particular metrics: Generalized Ellis-Bronnikov WH
 
 \begin{equation}
 	ds^2 = -dt^2 + dl^2 + r^2(l)[d\theta^2 + sin^2(\theta)d\phi^2]
 	\label{eqn15}
 \end{equation}

where

\begin{equation}
	r(l) = (b_0^m + l^m)^{\frac{1}{m}} 
	\label{eqn16}
\end{equation}
 with \textit{$m$} taking only even values to ensure smoothness of \textit{$r(l)$} over the entire domain. \textit{$\theta$} and \textit{$\phi$} are spherical coordinates, \textit{$l \in (-\infty, \infty)$} is the proper radial  coordinate and \textit{$b_0$} denotes the radius of a throat of the WH. Henceforth we use notations $c = 1$.
 
Examining a scenario when there is an accreting matter inside the WH (Ref. \refcite{Harko}) and by taking strong magnetic field into account (Ref. \refcite{Kardashev}), it is clear that any charged particle entering such an ambient will very soon lose the perpendicular momentum via the strong synchrotron emission process. As a result the particles will transit to the ground Landau level and will slide along the co-rotating field lines. Generally speaking, the particles which escape the regions of their origin, become force-free. From very general considerations, one can show that the only trajectory, that enables such a behaviour is the Archimede's spiral. Significance of such a rotating field line is that it leads to the possibility of achieving the terminal velocity as the radial coordinate approaches infinity (Ref. \refcite{AR GD ZO}). We examin conical spirals having the property of the Archimede's spiral 
 \begin{equation}
	\theta = const  \qquad \text{and} \qquad \Phi = \alpha l + \omega t,
	\label{eqn17}
\end{equation}

We see that \textit{$d \Phi = 0$} when the radial velocity equals \textit{$-\frac{\omega}{\alpha}$}. This means that in the laboratory frame of reference the angle of rotation does not change and therefore, the trajectory will be a straight line. We can analyse such a kinematical picture by means of the effective angular velocity, $\Omega\equiv d\Phi/dt = \omega+\alpha v$, thus the angular velocity of the particle seen by the laboratory observer. In particular, when the radial velocity equals the terminal velocity, the effective angular velocity vanishes, leading to the rectilinear trajectories in the laboratory frame of reference. It is worth noting that for a particle to have terminal velocity less than the speed of light, the field line can not have an arbitrary shape and the following condition has to be satisfied: \textit{$|\omega|<|\alpha|$}. However, if perchance this condition is violated, as velocity gets closer to one that of light, i.e. gamma factor increases, we would expect the charged particle to become energetic enough to start bending the field lines, since it is physically impossible for absolutely rigid rotations to exist. However, we will be examining cases when this criterion is well met, thus considering rotation of trajectories rigid.

Let us further examine restrictions that are imposed on moving the particle along such rotating spiral field lines with parameterization given by Eq.~(\ref{eqn17}). We consider total velocity at any given time, not just near its terminal value

\begin{equation}
	v_{tot}^2 = g_{ll} \biggl(\frac{d^2l}{dt^2}\biggr)^2 + g_{\theta \theta} \biggl(\frac{d^2\theta}{dt^2}\biggr)^2 + g_{\Phi\Phi} \biggl(\frac{d^2\Phi}{dt^2}\biggr)^2.
	\label{eqn18}
\end{equation}
By taking the fundamental principle of general relativity ($v_{tot}<1$ for massive particles) and Eqs. ~(\ref{eqn17},\ref{eqn18}) into account, we arrive at

\begin{equation}
	v^2 + sin^2(\theta)(b_0^m + l^m)^{\frac{2}{m}}(\alpha v + \omega)^2< 1.
	\label{eqn19}
\end{equation}
 
\begin{figure}[h]
	\centering
	\includegraphics[width=0.45\textwidth]{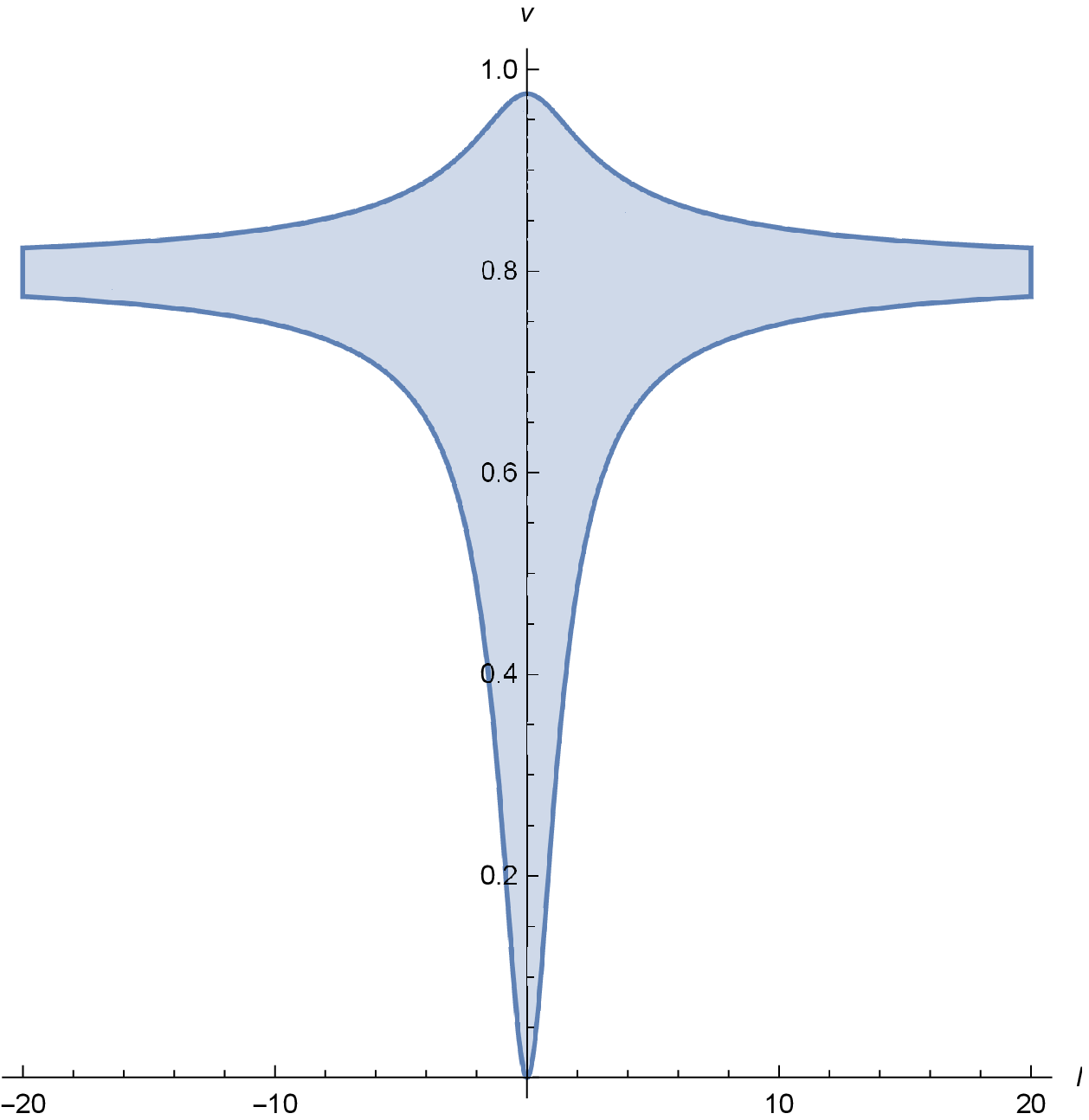}
	\includegraphics[width=0.45\textwidth]{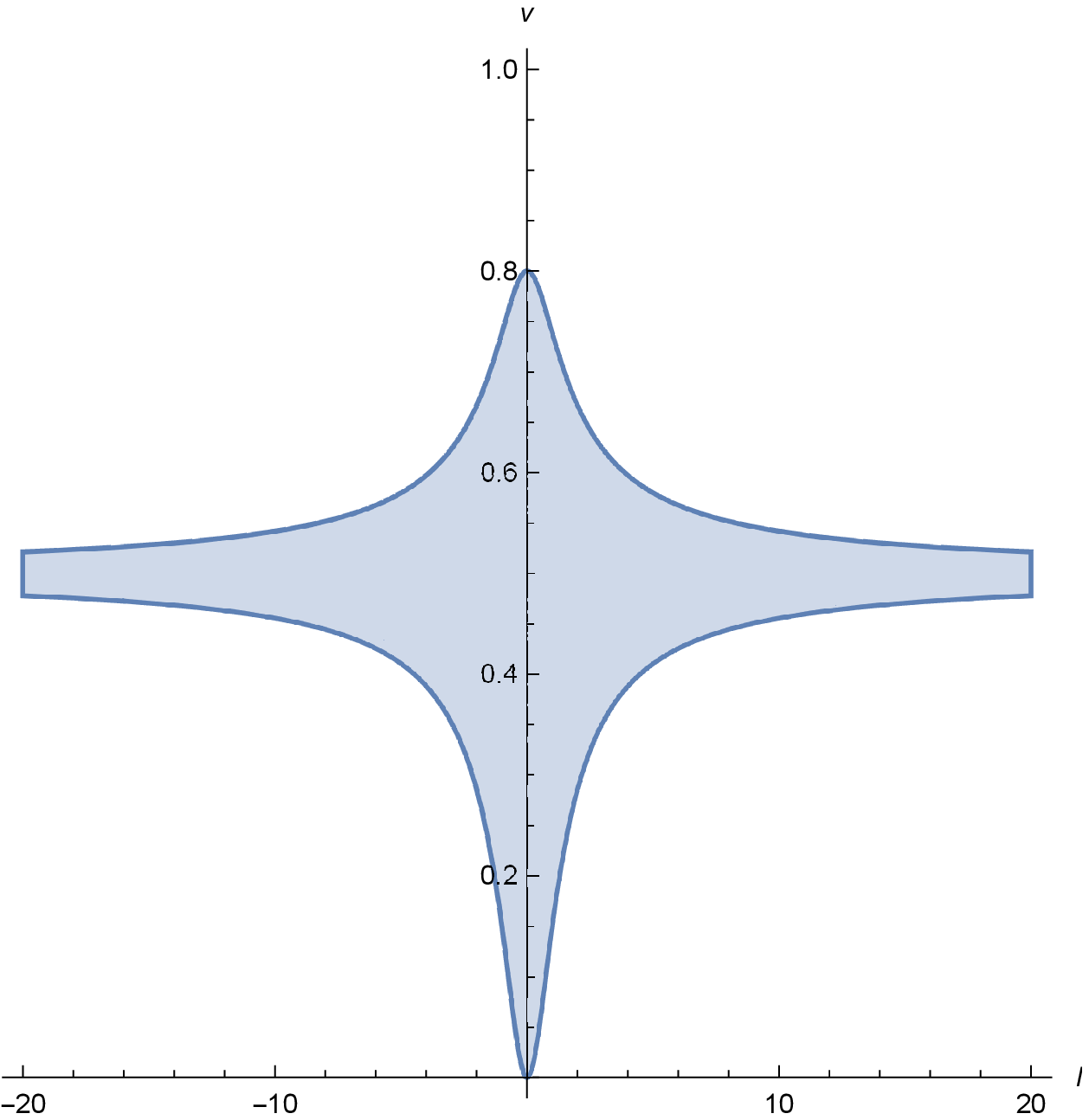}
	
	\caption{Allowed pairs of the radial velocity and the proper coordinate in the phase space for different $\alpha$. $\omega = 1$, $b = 1$, $m=2$, $\theta=\pi/2$ with  $\alpha = -1.25$ (left panel) and $\alpha = -2$ (right panel). \label{f1}}
\end{figure}

Eq.~(\ref{eqn19}) imposes a restriction on trajectories in the phase space, namely that they should lie in the region shown in Fig.~\ref{f1} determined by the above equation. In other words pairs of initial radial velocity and coordinate can only be chosen from within those regions and motion that follows should completely lie in it as well. Note that the allowed region of motion is the function of the characteristics of the WH as well. Moreover, if we aim to derive the equations of motion for the radial coordinate, possible initial values of the the radial coordinate and velocity become important to describe any motion. Let us consider the possible maximum and minimum values of the initial radial velocities at $l_0 = 0$. From Eq.~(\ref{eqn19}) one obtains

\begin{equation}
	v^{max}_{min} = \frac{-sin^2(\theta)b_0^2\omega \alpha  \pm \sqrt{1 + sin^2(\theta)b_0^2(\alpha^2 - \omega^2)}}{1 + sin^2(\theta)b_0^2\alpha^2}.
	\label{eqn20}
\end{equation}

We see that since \textit{$|\omega|<|\alpha|$}, \textit{$v_{max}$} will always be less than one and any possible motion will be bounded by it. It is also evident from Fig.~\ref{f1} that as the proper radial coordinate increases, possible velocities should tend to the terminal value.

To examine the actual motion we plug in the parameterisation given by Eq. (\ref{eqn19}) into Eq. (\ref{eqn17}) leading to the following metric tensor  
\begin{gather}
	g_{\alpha \beta}
	=
	\begin{bmatrix}
		-1 + \sin^2(\theta)(b^m + l^m)^{\frac{2}{m}}\omega^2 &
		\sin^2(\theta)(b^m + l^m)^{\frac{2}{m}}\omega \alpha \\
		\sin^2(\theta)(b^m + l^m)^{\frac{2}{m}}\omega \alpha &
		1 + \sin^2(\theta)(b^m + l^m)^{\frac{2}{m}}\alpha^2  
	\end{bmatrix}
	\label{eqn21}
\end{gather}

Since the GEB WH metrics is asymptotically Minkowskian and for \textit{$|\omega|<|\alpha|$}, we expect the radial velocity to reach the terminal value. Such a behavior in dynamics can be verified by expanding Eq. ~(\ref{eqn14}) for \textit{$l>>b_0$}, leading to (the same expression as in (Ref.~\refcite{Ars}))
\begin{equation}
	v = -\frac{\omega}{\alpha} + \frac{1+E^2 \pm E\sqrt{\alpha^2 - \omega^2}}{\omega \sin^2(\theta)}\frac{1}{l^2}.
	\label{eqn22}
\end{equation}
As it is clear from the above equation as radial coordinate tends to infinity, the radial velocity approaches its terminal value determined solely by the properties of the field line to which it is constrained. The fact that the radial velocity tends to the terminal value at infinities ensures that the numerator in Eq. (\ref{eqn7}) tends to zero there. Since the numerator is a smooth function of coordinates with no singularities, acceleration will have finite value if the metrics is non-degenerate, i.e. determinant appearing in the denominator of the aforementioned equation is always nonzero. That is easily verified by considering the metric tensor given by 
\begin{equation}
	\Delta g = -\biggl(1 + sin^2(\theta)(\alpha^2 - \omega^2)(b_0^m + l^m)^{\frac{2}{m}}\biggr).
	\label{eqn23}
\end{equation}
Since \textit{$|\omega|<|\alpha|$} and \textit{$m$} only taking even values, we see that the determinant always has a non-zero value and hence, the equation of motion will have no singularities in the middle region of the motion either. 

\begin{figure}[pb]
	\centering
	\includegraphics[width=0.45\textwidth]{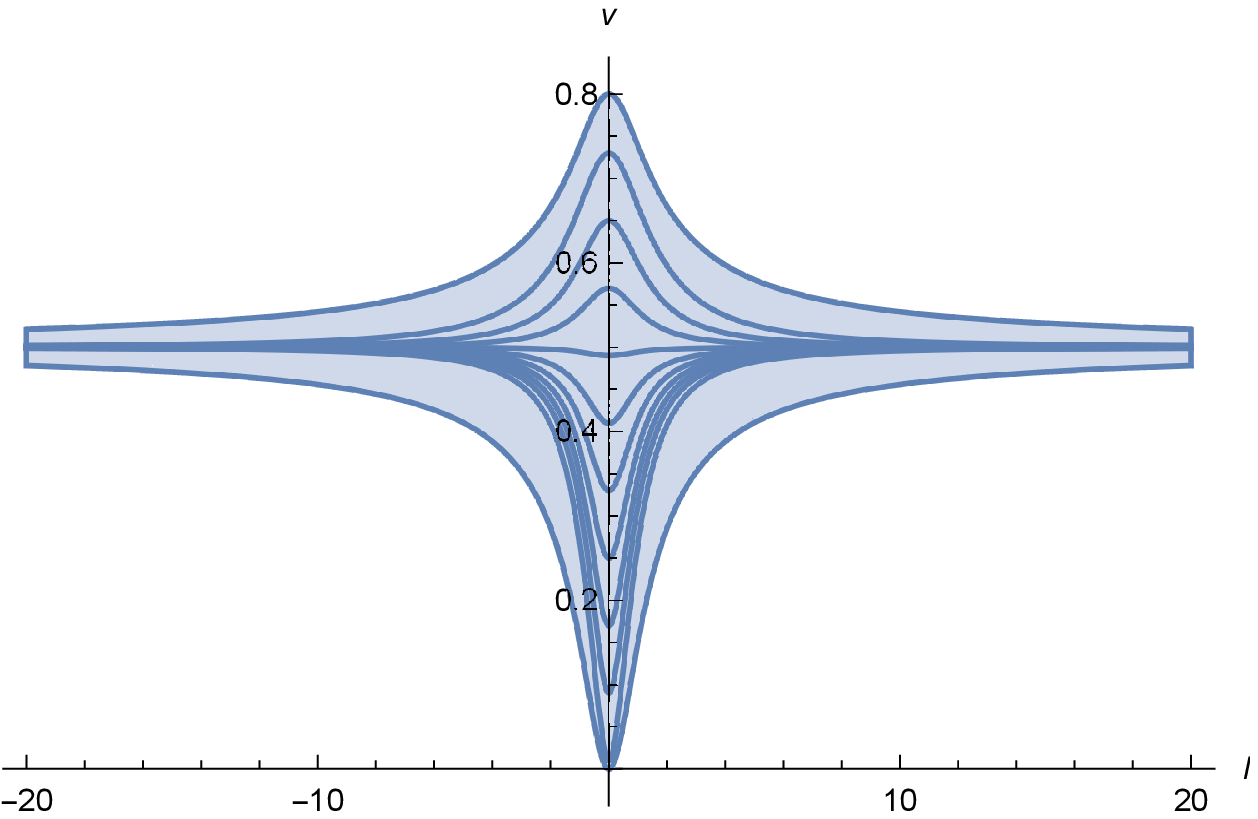}
	\includegraphics[width=0.45\textwidth]{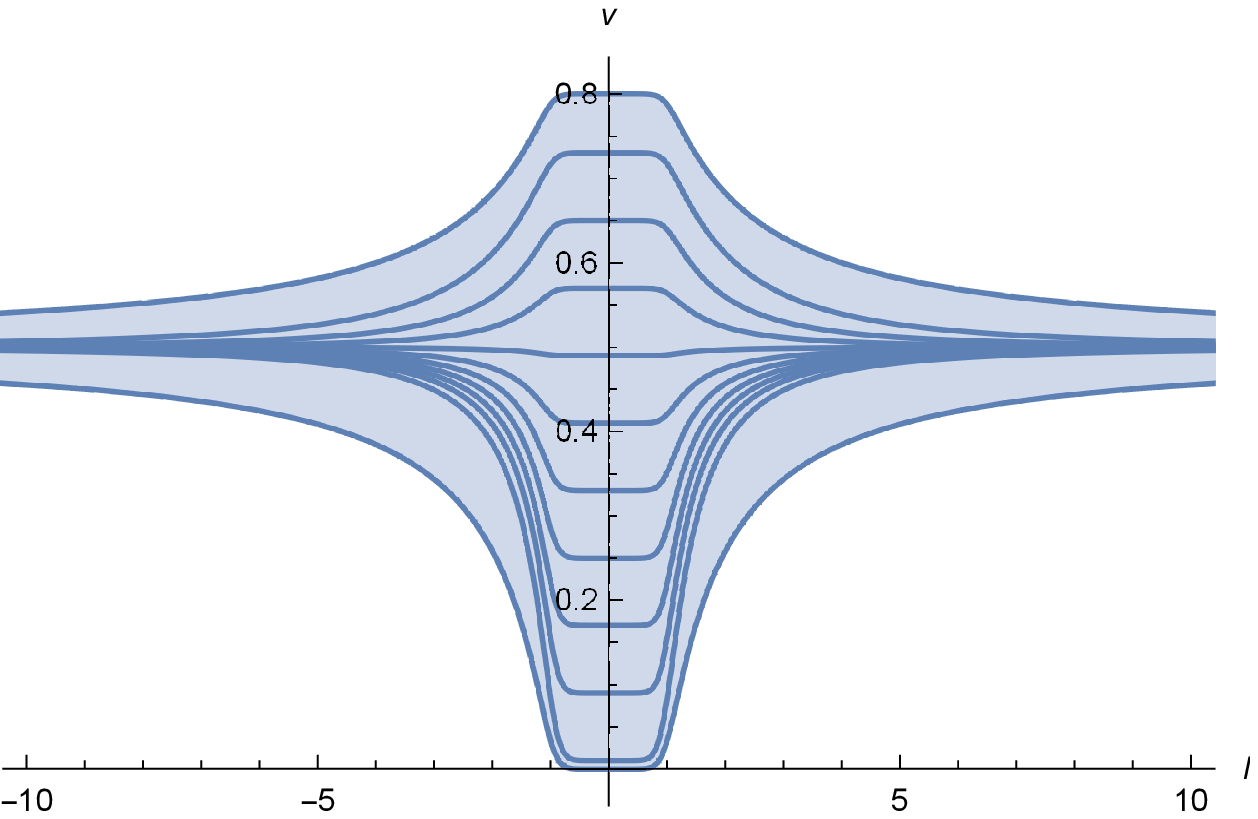}
	
	\vspace{\floatsep}
	
	\caption{Radial velocity dependence on the radial proper coordinate. The set of parameters is:	 $\omega = 1$, $\alpha = -2$, $b_0 = 1$, $\theta = \pi/2$ with $m = 2$ (left panel) and $m = 10$ (right panel). \label{f2}}
\end{figure}

Initial velocity fixes the conserved energy by Eq.~(\ref{eqn12}), that we choose to compute for \textit{$l_0=0$} and use it in Eq.~(\ref{eqn14}). In Fig.~\ref{f2} we see the trajectories of particles in the phase space for different values of the radial velocity at \textit{$l_0=0$}. On both panels the shaded area is the allowed region of motion given by Eq.~(\ref{eqn19}) and lines show the trajectories for different values of energy. It is seen that all trajectories lie within the aforementioned region for different initial velocities and different values of \textit{$m$} characterizing WH. 

Although we are using  \textit{$l_0=0$} and velocity there to calculate conserved energy (Eq.~(\ref{eqn14})), this method of solution with Eq.~(\ref{eqn14}) fixes all the points in the phase space with the given energy from  \textit{$l_0=-\infty$} to  \textit{$l_0=\infty$}. That is of particular interest since we are interested in particles entering the WH from one side region and then leaving it from the other.

From Fig.~\ref{f2} we see that while increasing the parameter \textit{$m$}, the particle undergoes acceleration further away from the origin. To study the dynamics of the particle in terms of the acceleration in a more detailed way we consider Eq. (\ref{eqn7}) for the metrics given by Eq. (\ref{eqn21}). The radial acceleration then, as a function of the radial coordinate and velocity, is easily derived

\begin{equation}
	\frac{d^2l}{dt^2} = -\frac{\sin^2(\theta)(\omega + \alpha v) (b_0^m + l^m)^{\frac{2}{m} - 1}l^{m-1}[\alpha v - \omega + 2\omega v^2 + \sin^2(\theta)(\omega + \alpha v)^2\omega r^2(l)]}
	{1+ \sin^2(\theta)(\alpha^2 - \omega^2)r^2(l)}
	\label{eqn24}
\end{equation}

As expected, when the radial velocity tends to its critical value, the acceleration vanishes due to the effective angular velocity \textit{$\Omega = \omega + \alpha v$}. Putting \textit{$m=2$} and thus reducing to the simplest form of GEB, we repeat the result derived in Ref~\refcite{Ars}. Therefore, we have found that vanishing of the acceleration when $\Omega = 0$ is a general property of the GEB metrics. 

\begin{figure}[pb]
	\centering
	\includegraphics[width=0.47\textwidth]{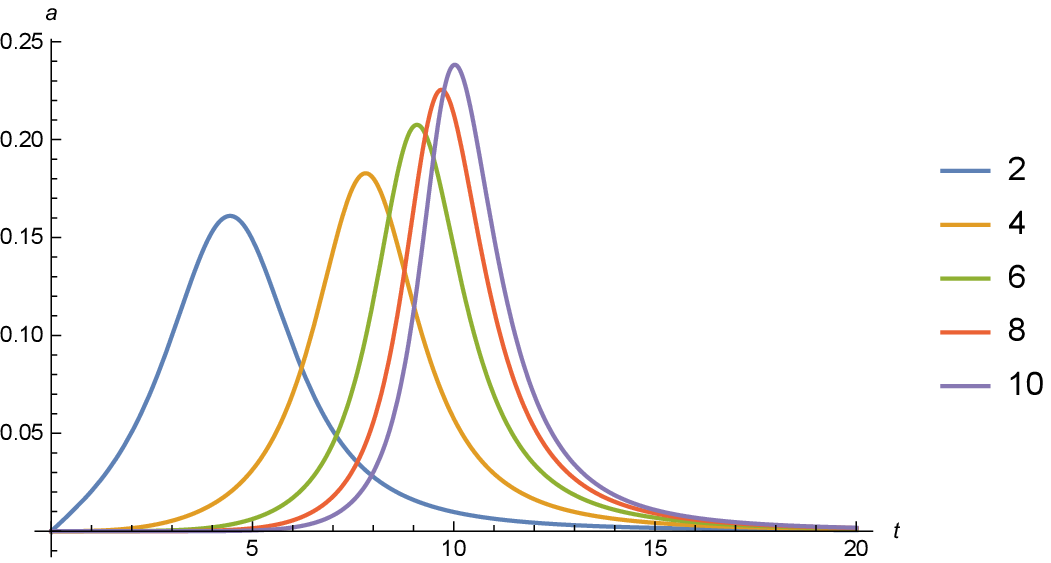}
	\includegraphics[width=0.47\textwidth]{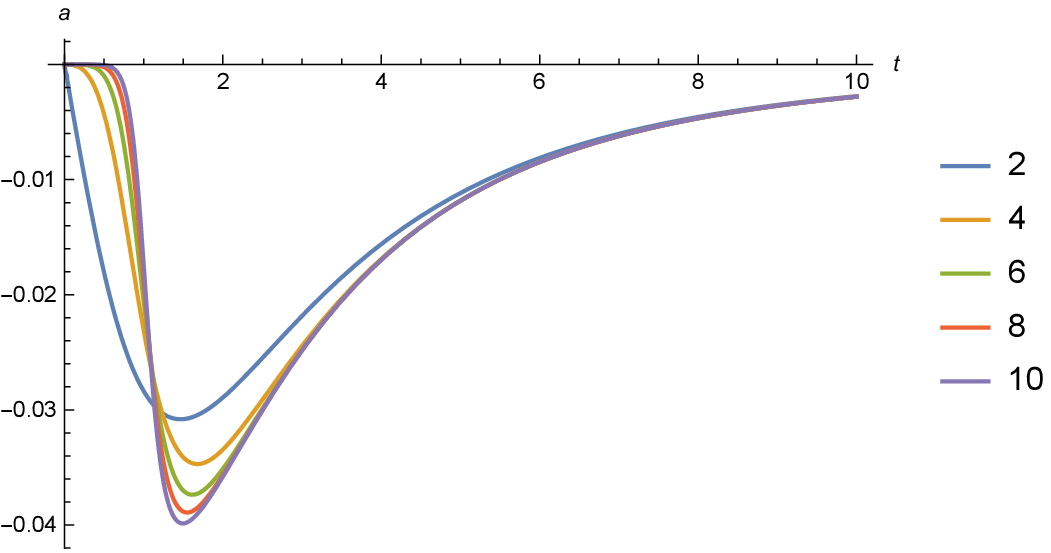}
	
	\vspace{\floatsep}
	
	\includegraphics[width=0.47\textwidth]{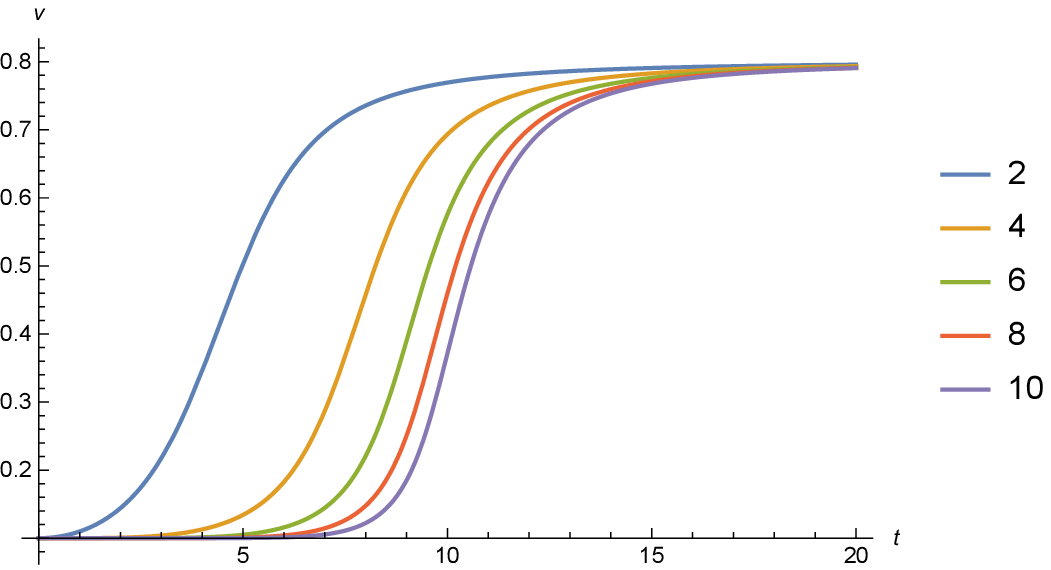}
	\includegraphics[width=0.47\textwidth]{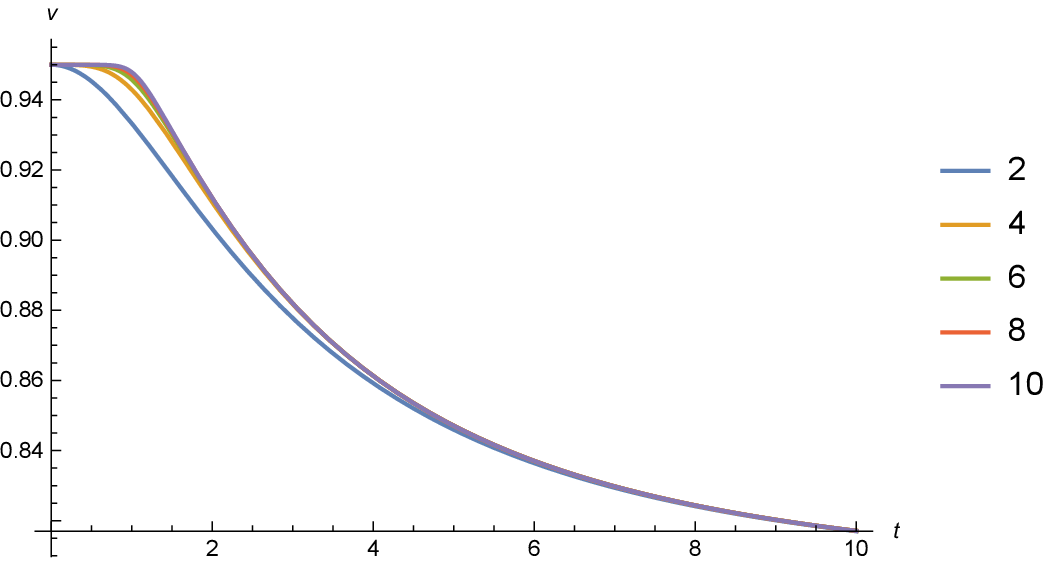}
	
	\vspace{\floatsep}
	\includegraphics[width=0.47\textwidth]{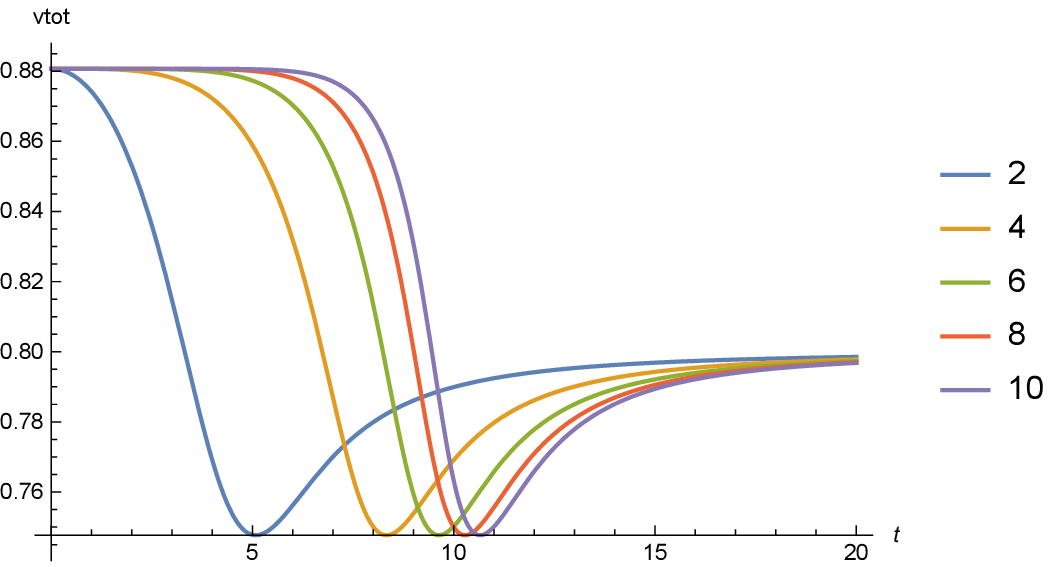}
	\includegraphics[width=0.47\textwidth]{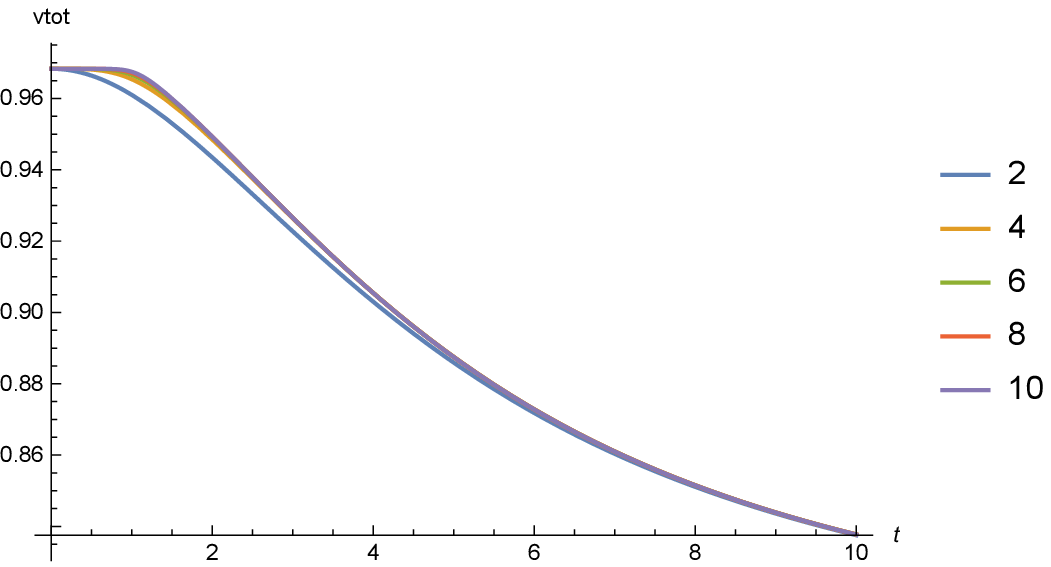}
	
	\caption{Radial acceleration (top panel), radial velocity (middle panel) and the total velocity (bottom panel) dependence on laboratory time. The set of parameters is: $\omega = 1$, $\alpha = -1.25$, $b_0 = 1$, $\theta = \pi/2$, $l_0 = 0$, $m = {2,4,6,8,10}$ with \textit{$v_0 = 0.1$} (left) and  \textit{$v_0 = 0.95$}  (right).  \label{f3}}
\end{figure}

From Eq. (\ref{eqn24}) one can see that for \textit{$l<<b_0$} and \textit{$m>>1$} the term \textit{$(b_0^m + l^m)^{\frac{2}{m} - 1}l^{m-1}$} is significantly reducing the value of acceleration, thus resulting in the initial plateaus in the right panel of Fig.~\ref{f2}. The same behaviour is evident graphically from Eq.(~\ref{eqn24}) that allows us to plot dependence of the radial acceleration on the laboratory time as shown in Fig.~\ref{f3} \footnote{It is worth noting that the corresponding figure in Ref.~\refcite{Ars} for $m = 0$ is not correctly plotted.}. It is clear that the bigger the value of parameter \textit{$m$}, later the particle experiences acceleration. It is evident from the top two graphs in Fig.~\ref{f3} that as the acceleration tends to zero, the radial velocity must reach the terminal value as shown in the bottom part of the Fig.~\ref{f3} for different values of \textit{$m$}.

\begin{figure}[pt]
	\centering
	\includegraphics[width=0.49\textwidth]{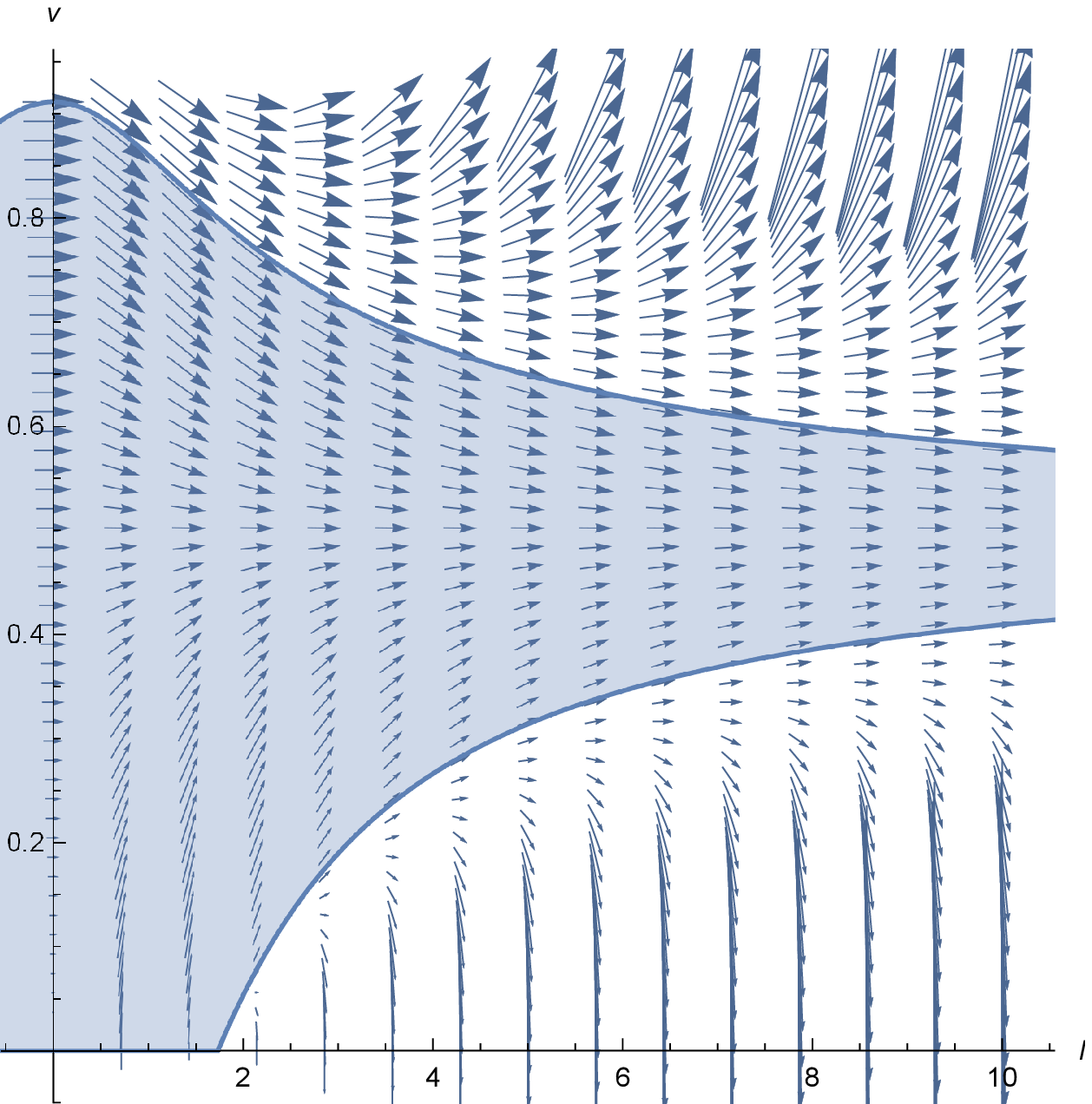}
	\includegraphics[width=0.49\textwidth]{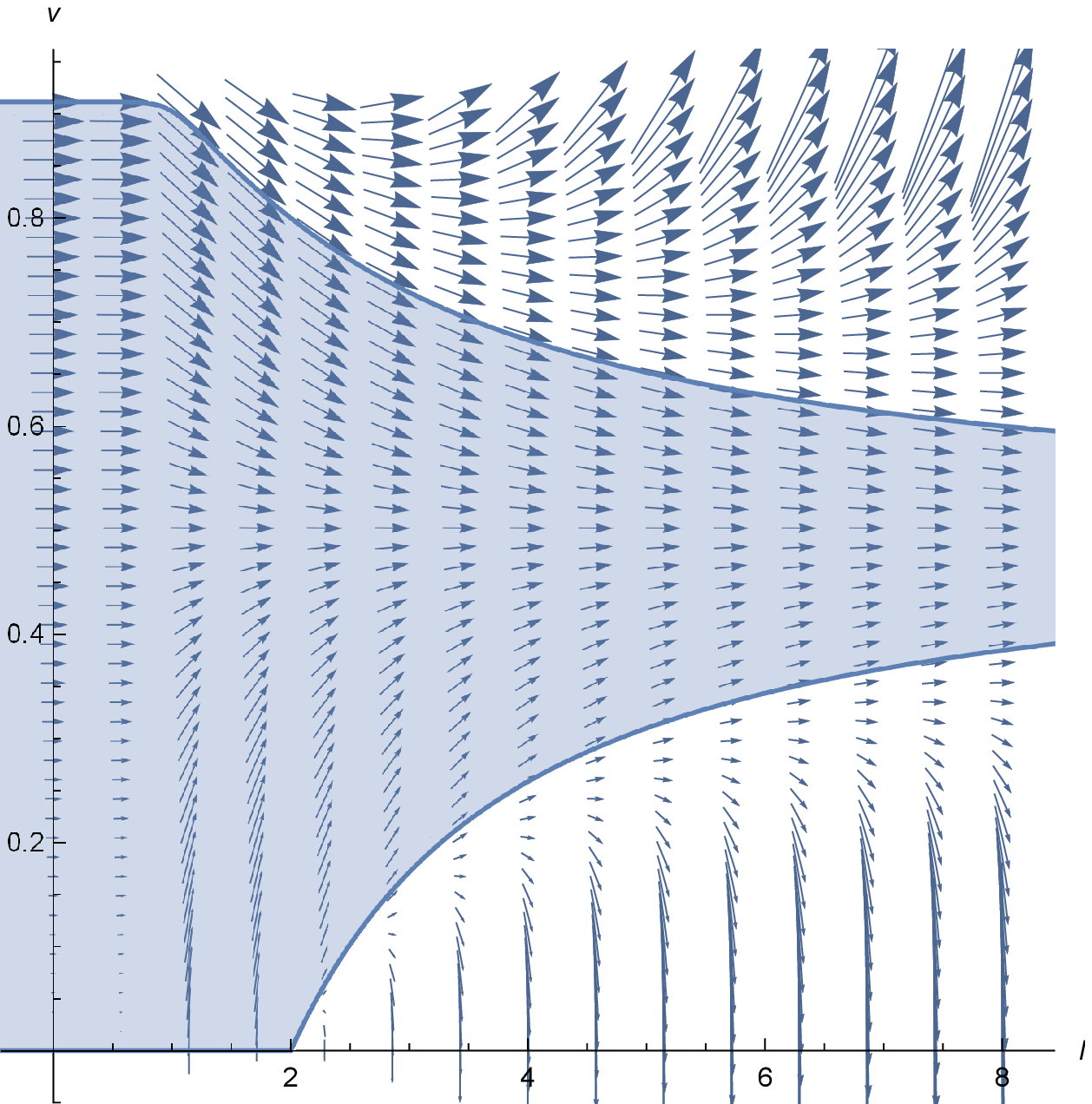}
	
		\caption{Vector field for the differential equation given by Eq. (\ref{eqn24}) The set of parameters is: $\omega = 1$, $\alpha = -2$, $b_0 = 1$, $\theta = \pi/6$, with $m = 2$ (left panel) and $m = 10$  (right panel).  \label{f4}}
\end{figure}

From Eq.~(\ref{eqn24}), we construct the vector field shown on Fig.~\ref{f4}. As it is evident, outside the shaded region, the solutions are unstable - arrows are pointing outwards and become vertical, which means that acceleration  changes very rapidly. 

\begin{figure}[pb]
	\centering
	\includegraphics[width=0.49\textwidth]{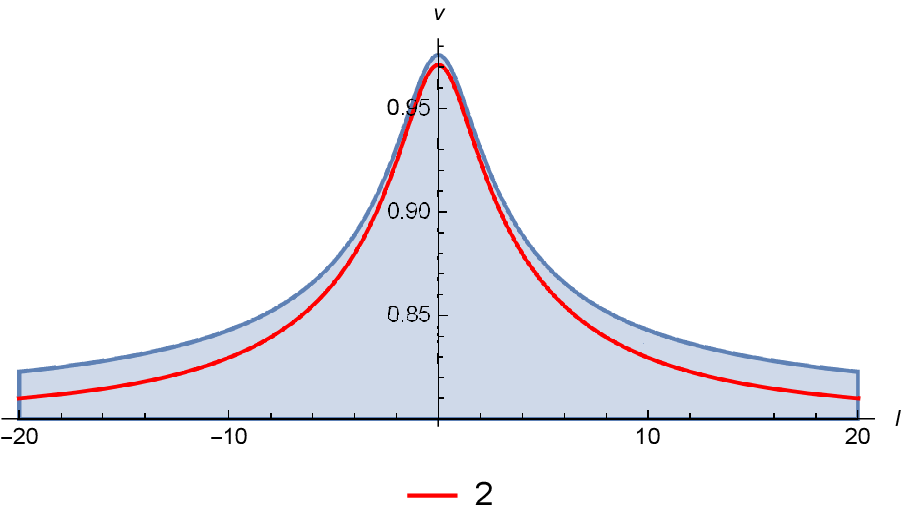}
	\includegraphics[width=0.49\textwidth]{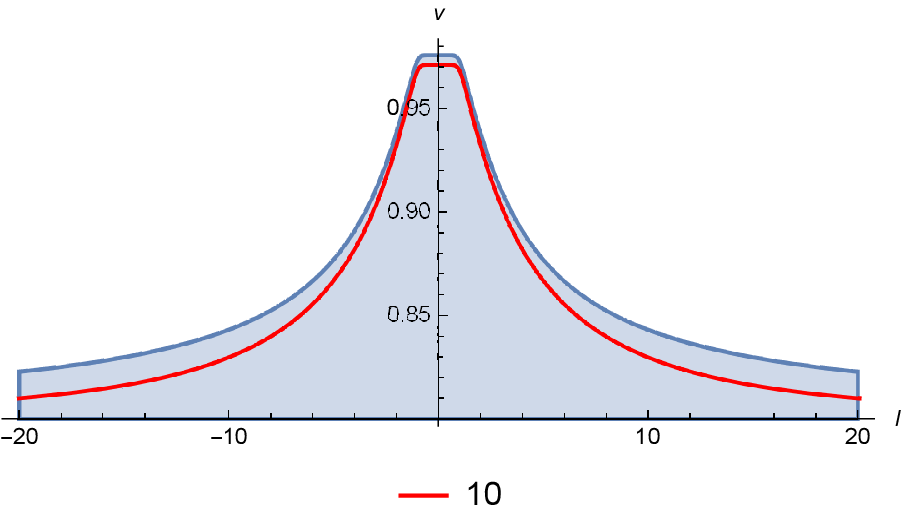}
	
	\caption{Radial velocity versus the radial coordinate from Eq.(~\ref{eqn24}). The set of parameters is: $\omega = 1$, $\alpha = -1.25$, $b_0 = 1$, $\theta = \pi/2$, with $m = 2$ (left panel) and $m = 10$ (right panel) \label{f5}}
\end{figure}

We saw from Fig.~\ref{f2} that for certain initial values of the radial velocity the particle may come from negative value of the initial radial coordinate, that is from the other side of the WH. In Fig.~\ref{f5} we see dependence of the radial velocity on the radial coordinate for two different values of $\textit{m}$. Note that the trajectory does not leave the allowed region of motion as it comes from  \textit{$l=-20$}, passes through the origin and then leaves WH.

\begin{figure}[pb]
	\centering
	\includegraphics[width=0.46\textwidth]{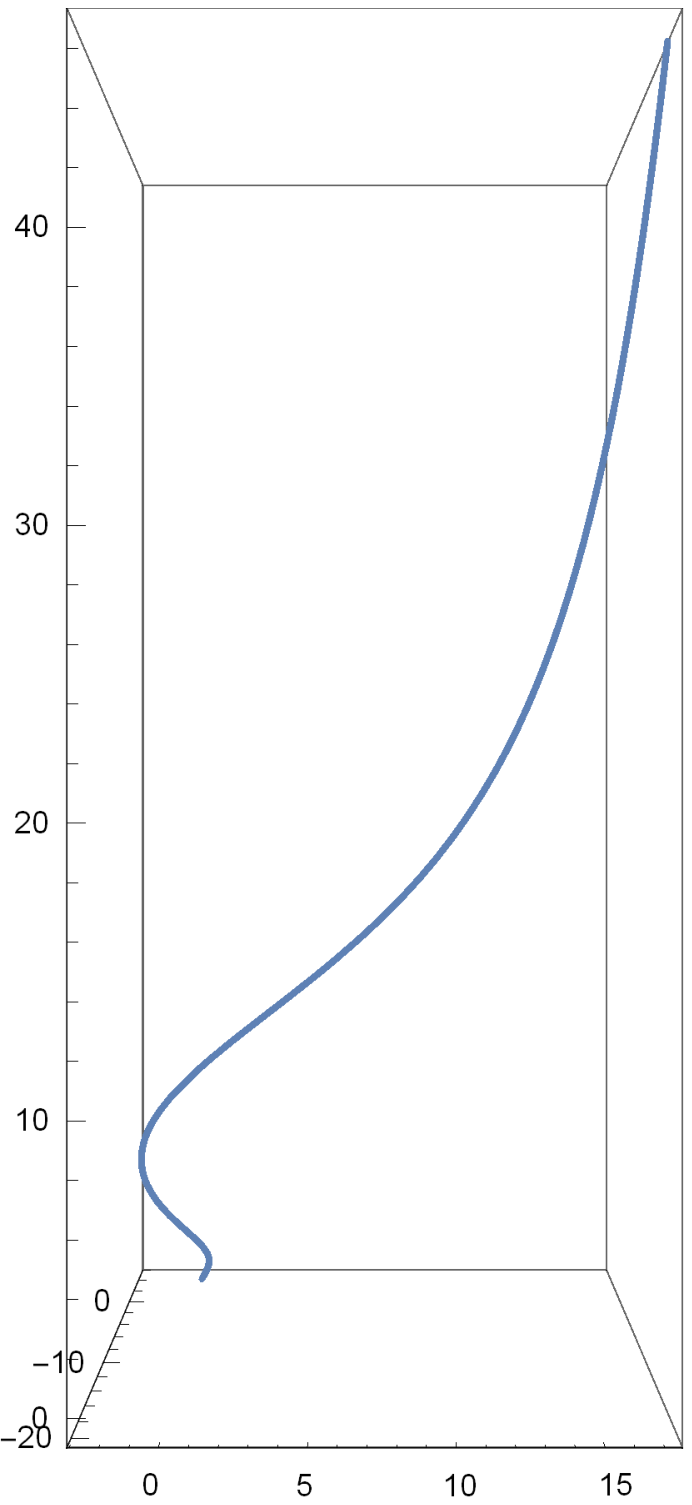}
	\includegraphics[width=0.49\textwidth]{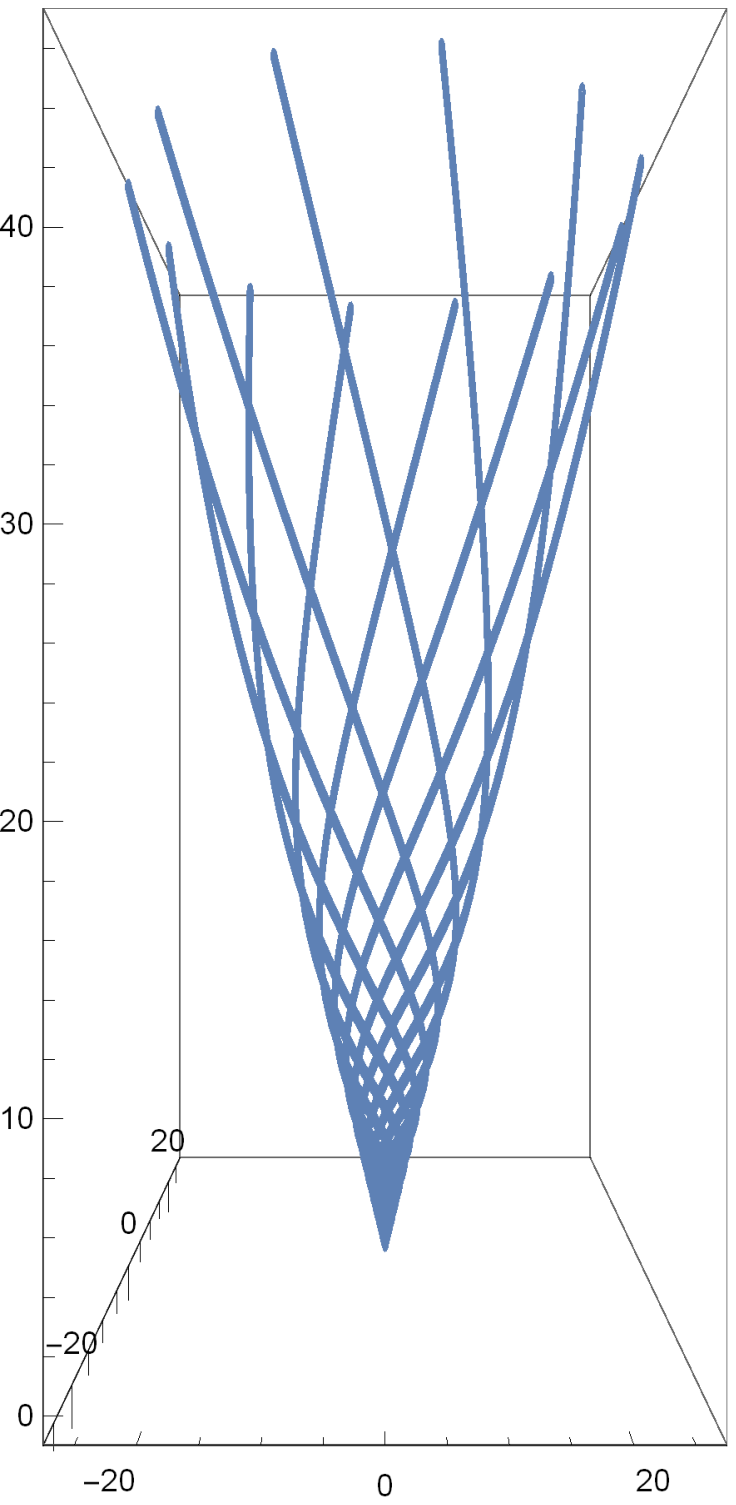}
	\caption{Trajectory of Motion of a single particle as observed from the laboratory frame of reference (left panel) and for the jet of particles (right panel). The set of parameters is: $\omega = 1$, $\alpha = -2$, $b_0 = 1$, $\theta = \pi/6$, $m = 2$, $l_0 = 0$ and $v_0=0.9$. \label{f6}}
\end{figure}

\begin{figure}[pt]
	\centering
	\includegraphics[width=0.48\textwidth]{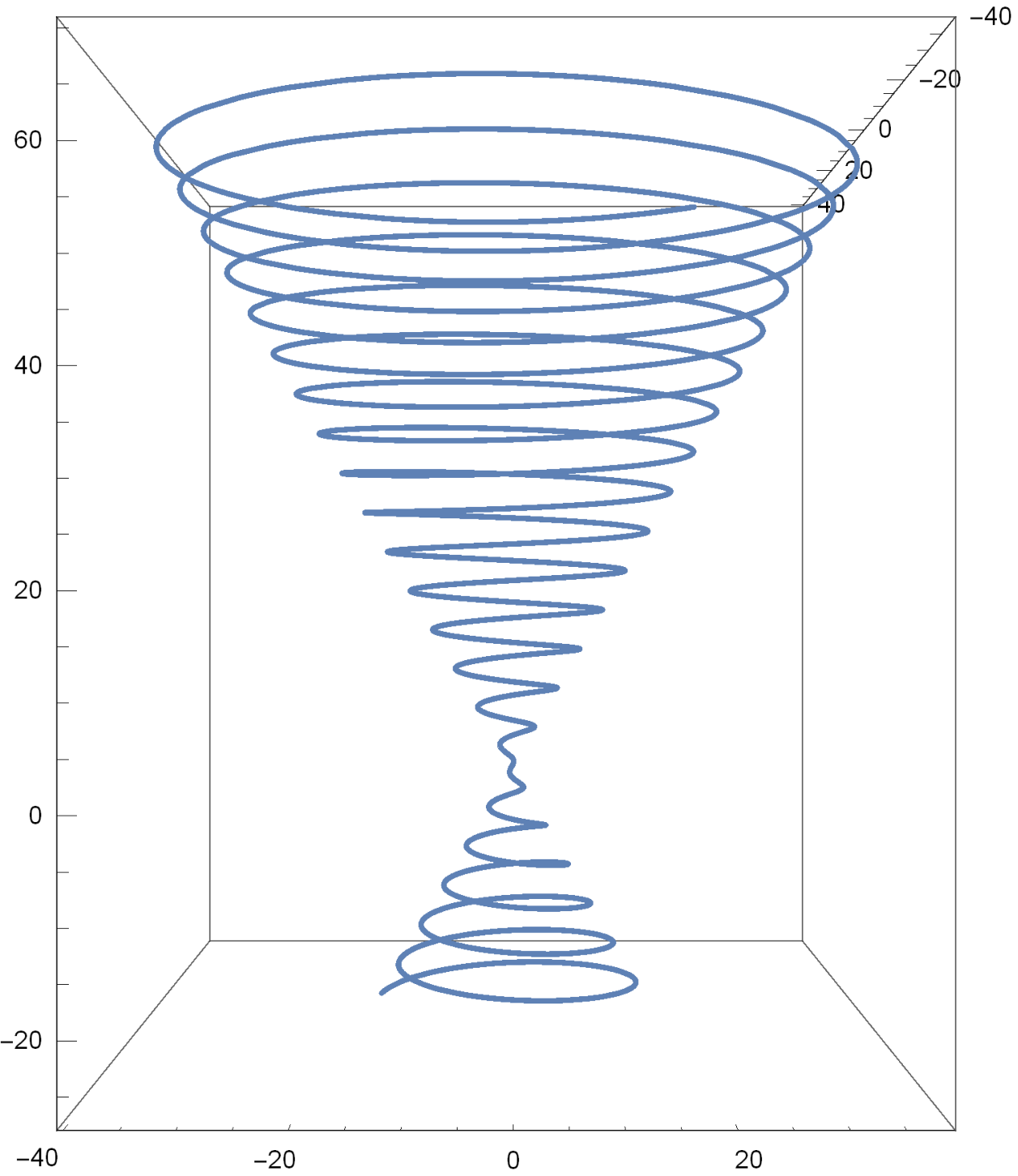}
	\includegraphics[width=0.50\textwidth]{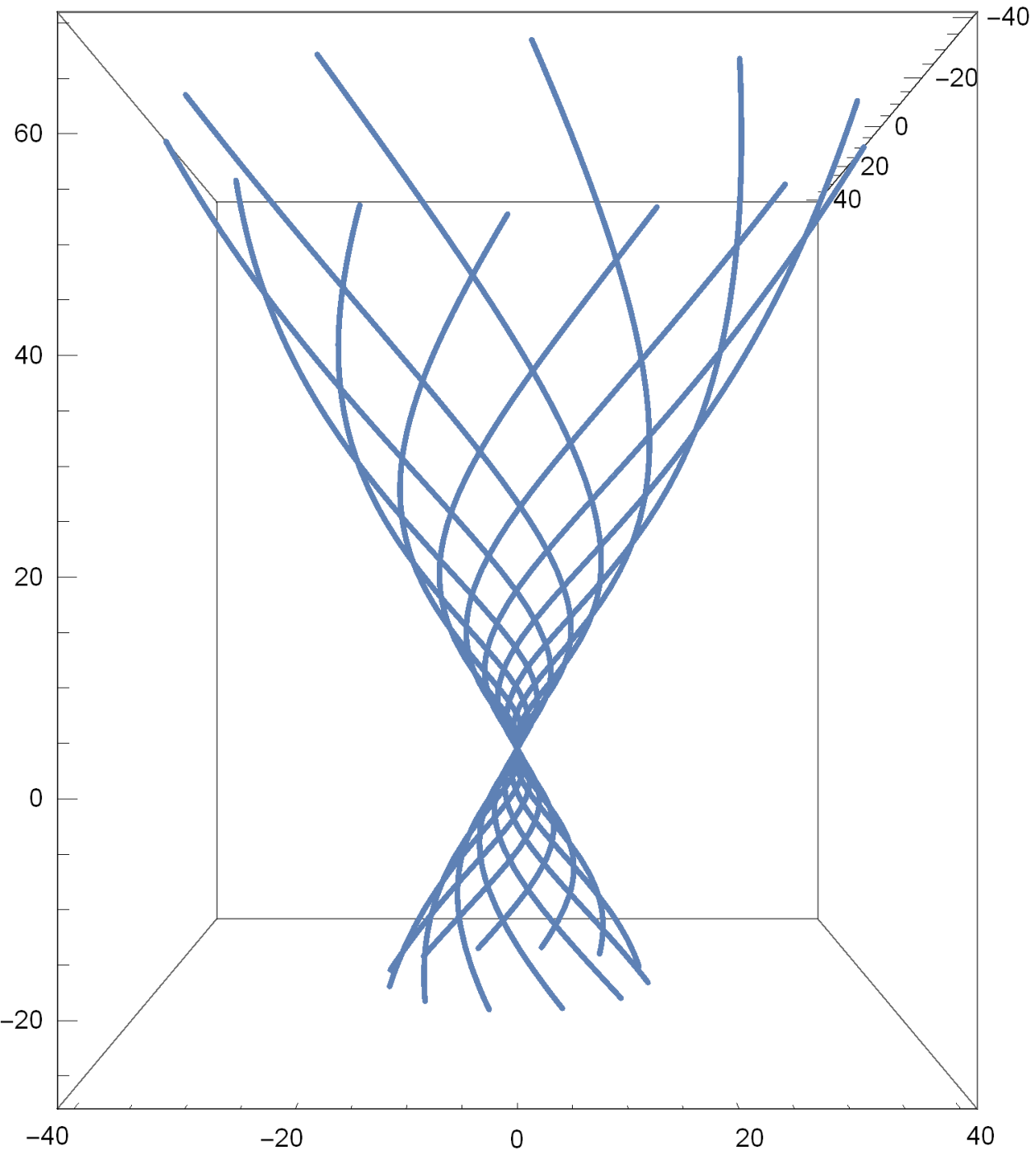}
	\caption{Trajectories of a single particle as seen from the local frame of reference (left panel) and the jet of particles observed from the laboratory frame of reference (right panel) $\omega = 1$, $\alpha = -1.25$, $b_0 = 1$, $\theta = \pi/6$, $m = 2$, $l_0 = -30$ and $v_0 = 0.827$.   \label{f7}}
\end{figure}

\begin{figure}[ph]
	\centering
	\includegraphics[width=0.49\textwidth]{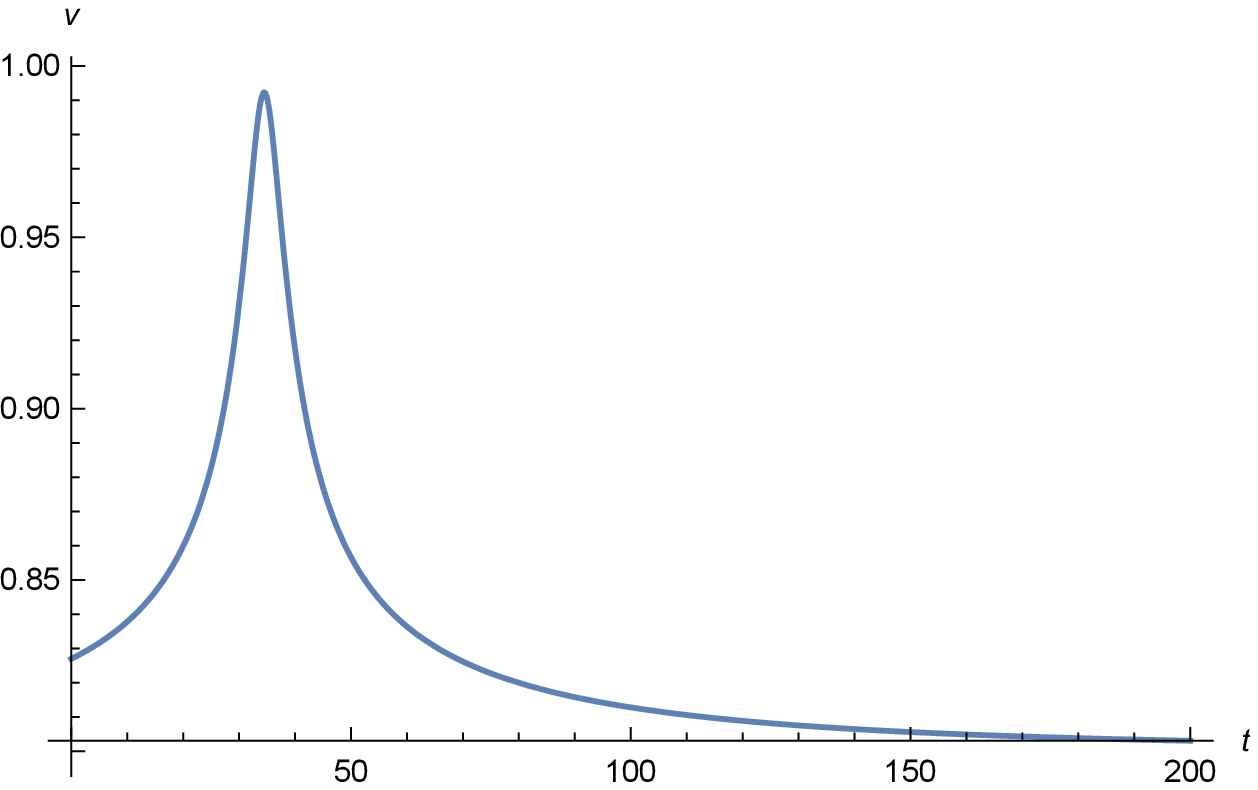}
	\includegraphics[width=0.49\textwidth]{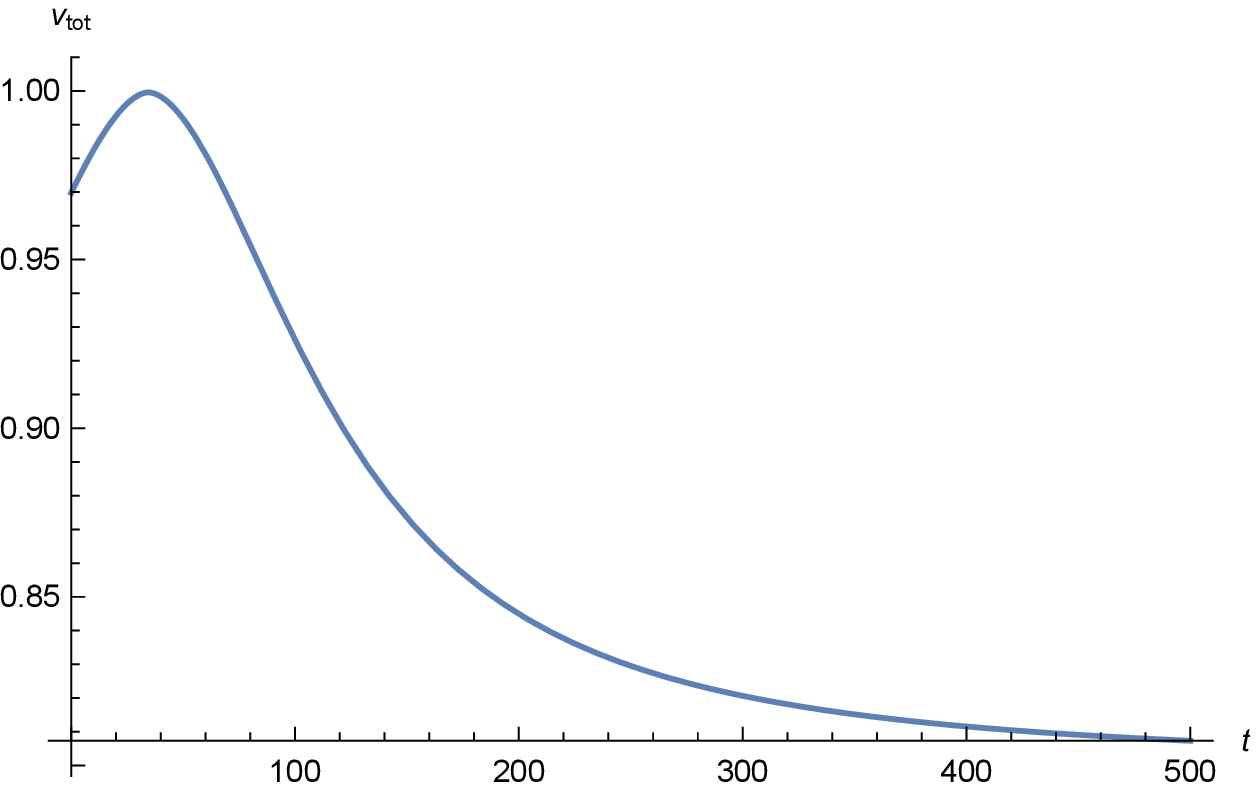}
	\caption{Radia velocityl (left panel) and total velocity (right panel) versus time. The set of parameters is: $\omega = 1$, $\alpha = -1.25$, $b_0 = 1$, $\theta = \pi/6$, $m = 2$. \label{f8}}
\end{figure}

In many astrophysical scenarios jets of ionised matter is emitted alongside the axis of rotation. In our case, since it is considered that in the central region of WHs some certain astrophysical objects might be located, we consider motion of the particle when $\theta<\pi/2$. In Fig.~\ref{f6} on the left panel we see the parametric plot for a position of a single particle as seen from the laboratory frame of reference and on the right panel we see how it would look like for the jet of numerous particles with phase shifts in the azimuth coordinate \textit{$\phi$}.

We see that in the beginning of the motion trajectory is curved, however, further from the origin, when the  force-free regime is achieved the particle slides on the spiral so that from the laboratory frame of reference its trajectory is rectilinear.

In Fig.~\ref{f7} we plot jets of particles entering the WH from the one side and then leaving it from the other. On the left panel we plot the trajectory of a single particle as seen from the local (rotating) frame of reference and on the right panel we show the jets of particles observed from the laboratory frame of reference. Here as well we notice that from laboratory frame of reference, we observe that, before particles enter and as they leave the intermediate region of WH their trajectory as observed form laboratory is linear.

For the same parameters as in Fig.~\ref{f7} we plot the radial velocity and the total velocity versus time in  Fig.~\ref{f8} and as it is evident, the particle's velocity remains less than the speed of light and at the end it tends to it's terminal value.




\section{Summary}

In this paper we considered the GEB WH and studied a single particle dynamics moving along a co-rotating magnetic field lines. For this purpose we rewrote the metrics for this particular case, reducing it to the $1+1$ scenario.

We have specified the characteristics of the phase space, the regions in which the motion is physically possible. By means of the energy method (conserved quantity) it has been shown that every motion lies within that area.

Later, having derived acceleration as a function of time in the laboratory frame of reference, we constructed the vector field. It has been shown that any motion lying outside the "allowed region" is unstable, thus not possible. It has been shown that particles may enter and then leave the WH, reaching the terminal velocity. 

Within this paper we aimed to investigate the motion of the particle in GEB WH and to show that under some special circumstances, such as choosing Archimede's spiral, the particles may enter and leave it. However, it is also interesting to study the same problem for different types of WHs. For future works to come we intend to examine different kinds of realistic field topologies for another WH metrics that might allow the force-free regime asymptotically to be established and the particle to escape.


\end{document}